\documentclass[
    final            
  ]
  {aipproc}

\layoutstyle{8x11double}

\newcommand{\lsim}{\raisebox{-0.13cm}{~\shortstack{$<$ \\[-0.07cm] $\sim$}}~} 
\newcommand{\gsim}{\raisebox{-0.13cm}{~\shortstack{$>$ \\[-0.07cm] $\sim$}}~}

\newcommand{\beq}{\begin{eqnarray}} 
\newcommand{\eeq}{\end{eqnarray}} 

\begin{document}

\title{The constrained NMSSM: mSUGRA and GMSB 
\footnote{Talk given at SUSY~08, Seoul, Korea, June 16-21, 2008}
}


\author{Ulrich Ellwanger}
  {address={Laboratoire de Physique Th\'eorique\footnote{Unit\'e mixte
de Recherche -- CNRS -- UMR 8627} \\
Universit\'e de Paris XI, F-91405 Orsay Cedex, France}
}

\begin{abstract}
We review different constrained versions of the NMSSM: the fully
constrained cNMSSM with universal boundary conditions for gauginos and
all soft scalar masses and trilinear couplings, and the NMSSM with soft
terms from Gauge Mediated Supersymmetry Breaking. Regarding the fully
constrained cNMSSM, after imposing LEP constraints and the correct dark
matter relic density, one single parameter is sufficient to describe the
entire Higgs and sparticle spectrum of the model, which then contains
always a singlino LSP. The NMSSM with soft terms from GMSB is
phenomenologically viable if (and only if) the singlet is allowed to
couple directly to the messenger sector; then various ranges in
parameter space satisfy constraints from colliders and precision
observables. Motivations for and phenomenological features of extra
$U(1)'$ gauge symmetries are briefly reviewed.

\end{abstract}

\maketitle


\section{Introduction}

The Next-to-Minimal Supersymmetric Standard~Model (NMSSM) \cite{nmssm}
solves in a natural and elegant way the so-called $\mu$-problem
\cite{kim} of the MSSM: Within any super\-symmetric (SUSY) extension of
the Standard~Model (SM), a supersymmetric Higgs(ino) mass term $|\mu|
\gsim$ $100$~GeV is necessary in order to satisfy the LEP constraints on
chargino masses, but $|\mu| \lsim M_{SUSY}$ is required in order that
the effective potential develops a non-trivial minimum with 
$\left<H_u\right>$, $\left<H_d\right> \neq 0$. (Here $M_{SUSY}$ denotes
the order of magnitude of the soft SUSY breaking scalar masses as
$m_{H_u}$ and $m_{H_d}$.) The question is, why a supersymmetric mass
parameter as $\mu$ happens to be of the same order as $M_{SUSY}$.

In the NMSSM, an (effective) $\mu$-term is generated by the vacuum
expectation value (VEV) of an
additional gauge singlet superfield $S$ and a corresponding Yukawa
coupling, similarly to the way how quark and lepton masses are generated
in the SM by the VEV of a Higgs field. To this end, the
$\mu$-term in the superpotential $W$ of the MSSM,
$ W_{MSSM} = {\mu} H_u H_d\ +\dots\; $,
has to be replaced by
\beq
W_{NMSSM} = {\lambda} S H_u H_d 
+\frac{1}{3}{\kappa}S^3\ +\dots
\eeq
and the soft SUSY breaking term $ {\mu B} H_u H_d $ by
\beq
{\lambda A_\lambda} S H_u H_d  +\frac{1}{3} {\kappa A_\kappa} S^3\; .
\eeq

Assuming that all soft SUSY breaking terms are of ${\cal{O}}(M_{SUSY}$),
one obtains $\left< S\right> \sim M_{SUSY}/\kappa$ and hence an
effective $\mu$-parameter $\mu_{eff} \equiv \lambda \left< S\right> \sim
{\lambda}/{\kappa}\ M_{SUSY}$, which is of the desired order if 
${\lambda}/{\kappa} \sim {\cal{O}}(1)$. Instead of the two parameters
$\mu$ and $B$ of the MSSM, the NMSSM contains four parameters $\lambda,\
\kappa,\ A_\lambda$ and $A_\kappa$, and the spectrum includes one
additional CP-even Higgs scalar, one CP-odd Higgs scalar and one
additional neutralino from the superfield $S$. Generally, these states
mix with the Higgs scalars and neutralinos of the MSSM. Then, each of
the neutralino/CP-even/CP-odd sectors can give rise to a phenomenology
different from that of the MSSM:

a) The Lightest Supersymmetric Particle (LSP) can be dominantly
singlino-like (consistent with WMAP constraints on $\Omega h^2$
\cite{wmap}, if its mass is only a few GeV below the one of the
Next-to-LSP (NLSP), see \cite{belsemenov} and below) implying an
additional contribution to sparticle decay chains; note that the NLSP
could have a long life time leading to observable displaced vertices
\cite{singdecay};

b) The SM-like CP-even Higgs scalar $h_1$ can be $\sim
15$~GeV heavier than in the MSSM (at {low} $\tan\beta$!);

c) A CP-odd Higgs scalar $a_1$ can be (very) light (see also the talk by
J.~Gunion, these proceedings). A light CP-odd Higgs scalar can have an
important impact on B physics (see the talk by M. Sanchis-Lozano, these
proceedings), and can imply that the lightest CP-even scalar $h_1$
decays dominantly into $h_1 \to a_1\, a_1$ \cite{raxion,lighta1}. Then,
LEP constraints on $h_1$ are less restrictive, but the search for $h_1$
at the LHC can become considerably more difficult.

Note that these are not ``unavoidable'' predictions of the NMSSM, but
depend on the unknown parameters $\lambda$, $\kappa$, $A_\lambda$,
$A_\kappa$, $\tan\beta$ and $\mu_{eff}$. In the following we
investigate, amongst others, the phenomenological consequences of
particular boundary conditions on the parameters of the NMSSM at a high
scale like mSUGRA (universaly boundary conditions for gauginos and all
soft scalar masses and trilinear couplings at the GUT scale), and GMSB
(Gauge Mediated Supersymmetry Breaking).

The subsequent results are obtained with the help of the Fortran code
{\sf NMHDECAY/NMSSMTools}~\cite{nmssmtools}, which computes the Higgs
and sparticle spectra and Higgs branching ratios including radiative
corrections for general/mSUGRA/GMSB boundary conditions, and checks for
constraints from colliders/B-physics/(g-2)$_\mu$/\break dark matter (the
latter via MicrOMEGAs  \cite{belsemenov}).

\section{The cNMSSM}
By definition, the soft SUSY breaking gaugino, scalar masses and
trilinear couplings in the fully constrained cNMSSM -- including the
singlet sector -- are assumed to be universal (equal to $m_0$, $M_{1/2}$
and $A_0$, respectively) at the scale $M_{GUT} \sim M_{Planck}$ 
as generated via mSUGRA, i.e. minimal supergravity with
flavour-blind kinetic functions
\footnote{The results of this section have been obtained in
collaboration with A. Djouadi and A. M. Teixeira in \cite{univ}.}. As a
result, the number of unknown parameters is reduced to 4. In the
convention where $\kappa$ is implicitly determined by $M_Z$, these can
be chosen as $M_{1/2}$, $m_0$, $A_0$ and $\lambda$; one of these
parameters can still be replaced by $\tan\beta$. (A slightly less
constrained version of the cNMSSM, where the SUSY breaking mass $m_S$ of
the singlet is allowed to differ from $m_0$, has recently been studied
in \cite{balazs}; see also the talk by C. Bal\'azs, these proceedings.)

First, it is useful to recall the constraints on these parameters which
follow from a stable real (in order to avoid problems with CP-violating
observables) VEV of $S$ \cite{genNMSSM2}: the numerically most relevant
terms in the $S$-dependent part of the potential are
\beq
V(S) \sim m_S^2|S|^2 + \frac{\kappa}{3} A_\kappa (S^3 + S^{*3})
+\kappa^2 |S|^4 + \dots\; ,
\eeq
hence $V(S)$ has a stable nontrivial minimum only if $m_S^2 < {1}/{9}\
A_\kappa^2$, where $\kappa A_\kappa \left<S\right> < 0$. Since the
parameters $m_S$ and $A_\kappa$ are hardly renormalized between
$M_{GUT}$ and $M_{SUSY}$ (and $\kappa \left<S\right> > 0$ for $\mu_{eff}
>0$, which is desired for the correct anomalous magnetic moment of the
muon), one obtains the approximate inequalities
\beq\label{eq:m0a0bounds}
m_0 \lsim \frac{1}{3}|A_0|,\ A_0 < 0\; .
\eeq

Additional constraints follow from the properties of the LSP (which will
be the constituent of the dark matter), and the WMAP result \cite{wmap}
on the dark matter relic density:

First, for small values of $m_0$ (as the ones required by 
(\ref{eq:m0a0bounds})), the lightest stau $\tilde{\tau}_1$ would be the
LSP in the MSSM, which would be unacceptable due to its electric charge.
In the NMSSM, the additional (singlet-like) neutralino $\tilde{\chi}_1$ 
(with a mass proportional to $|A_\kappa| \sim |A_0|$) is lighter than
the $\tilde{\tau}_1$, if $|A_0|$ satisfies $|A_0| \lsim
\frac{1}{3}M_{1/2}$. Then (\ref{eq:m0a0bounds}) gives
\beq\label{eq:m0bound}
\ m_0 \lsim \frac{1}{10}M_{1/2}\; ,
\eeq
which would lead to an unacceptable LSP within the MSSM.

Second, in order to allow for a sufficiently rapid $\tilde{\chi}_1$
annihilation in the early universe (such that its relic density complies
with WMAP constraints), the $\tilde{\chi}_1-\tilde{\tau}_1$ mass
difference must be relatively small $(m_{\tilde{\tau}_1} -
m_{\tilde{\chi}_1} \sim  (1-8)$~GeV), and both masses must not be too
large (below $\sim 600$~GeV). Together, these constraints imply
\beq\label{eq:a0m12bounds}
A_0 \sim -\frac{1}{4}M_{1/2},\ M_{1/2} \lsim 2-3\ \mathrm{TeV}\; .
\eeq

Finally, the lower bound of $\sim 100$~GeV on $m_{\tilde{\tau}_1}$ from
LEP requires
\beq\label{eq:m12bound}
M_{1/2} \gsim 400\ \mathrm{GeV}\; .
\eeq

Then we find that, for $\lambda$ small enough (see below), the SM-like
Higgs scalar $H_{SM}$ has a mass $m_{H_{SM}} = 115 - 120$~GeV
(increasing with $M_{1/2}$) in agreement with LEP constraints. However,
for larger $\lambda$ the mixing of $H_{SM}$ with the singlet-like scalar
increases leading to a decrease of its mass $m_{H_{SM}}$. Hence
$\lambda$ must be relatively small,
\beq\label{eq:lbound}
\lambda \lsim 2\times 10^{-2}\; .
\eeq
(The NMSSM specific positive contribution to $m_{H_{SM}}^2$ proportional
to $\lambda^2$ \cite{nmssm} is negligible here, since $\tan\beta$ turns
out to be fairly large, see below.)

Hence, from (\ref{eq:m0bound}) and (\ref{eq:lbound}), neither $m_0$ nor
$\lambda$ have an important effect on the Higgs- and sparticle spectrum;
$A_0$ being determined by (\ref{eq:a0m12bounds}), the spectrum is
practically completely fixed by $M_{1/2}$.

\begin{figure}[ht]
\includegraphics[width=0.69\linewidth,angle=-90]{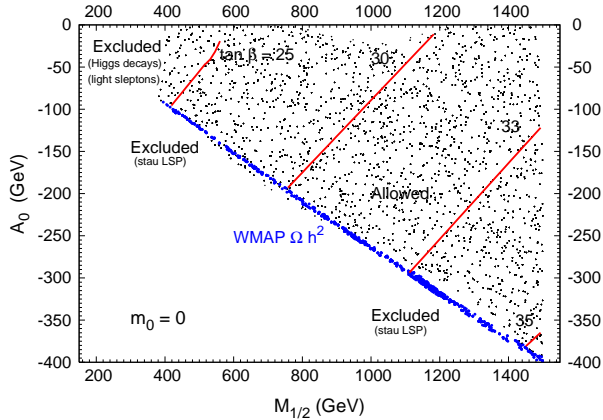}
\vspace*{-2mm}
\caption{The viable cNMSSM region in the $[M_{1/2},A_0]$  plane for $m_0
\sim  0$ and $\lambda=2 \times 10^{-3}$, once  theoretical, collider and
cosmological constraints have been imposed.}
\label{fig:1}
\end{figure}

In Fig.~\ref{fig:1} we show acceptable points in the $[M_{1/2},A_0]$
plane for $m_0 \sim 0$ and $\lambda=2 \times 10^{-3}$, which satisfy
theoretical and collider constraints; the blue line corresponds to the
additional satisfaction of WMAP constraints. For points above this line
the dark matter relic density comes out (far) too large. Also indicated
are lines of constant $\tan\beta$ (in red), which is seen to vary
between 25 and $\sim$~38 (for $M_{1/2}$ below 1.5~TeV as required for a
correct relic density for $m_0 \sim 0$).

\begin{figure}[!ht]
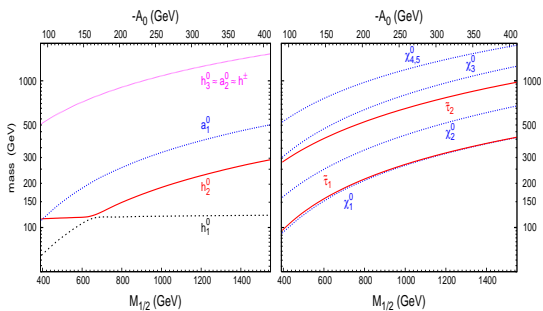

\mbox{
\includegraphics[width=4.cm,height=3.5cm,angle=-90]{Plot.phenoH.epsi} 
\includegraphics[width=4.cm,height=3.5cm,angle=-90]{Plot.phenoX0.epsi} 
}
\vspace*{-3mm}
\caption{The Higgs (left) and neutralino plus stau (right) mass spectra
in GeV  as a function of $M_{1/2}$ along the dark matter line; the
values of $A_0$ are   indicated in the upper axis.}
\label{fig:2}
\end{figure}

Still for $m_0 \sim 0$ and $\lambda=2 \times 10^{-3}$ (and $A_0$ along
the blue line in Fig.~\ref{fig:1}), we show in Figs.~\ref{fig:2} the
Higgs, neutralino and stau spectrum as function of $M_{1/2}$. The squark
and gluino masses are (except for the somewhat lighter stop masses)
of the order $2 \times M_{1/2}$.

Note that, for $M_{1/2} \lsim 640$~GeV, the lightest CP-even scalar
$h_1^0$ is singlet-like; however, due to the small value of $\lambda$,
its couplings to SM particles (as the $Z$-boson) are so
small that its mass is not constraint by LEP and, likewise, it will be
practically invisible at the LHC.

\begin{figure}[!ht]
\includegraphics[width=7cm]{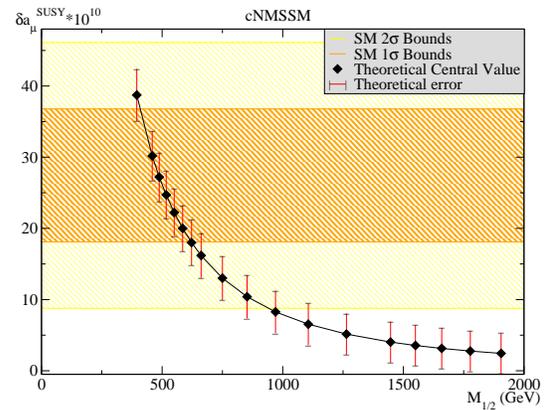}
\caption{$\delta a_\mu^{SUSY}$ as function of $M_{1/2}$ together with
the SM (experimental) $1\sigma$ and $2\sigma$ bounds.}
\label{fig:3}
\end{figure}

Actually, the parameter regions shown above satisfy all present
collider- and $B$-physics constraints, but do not necessarily describe
the deviation $\delta a_\mu$ of the anomalous magnetic moment $a_\mu =
(g_\mu - 2)/2$ from its SM value observed by the E821 experiment at BNL
\cite{bnl}. In \cite{Domingo:2008bb}, the dependency of $\delta a_\mu$
on $M_{1/2}$ (which is practically independent from $m_0$ and $\lambda$)
has been studied with the result shown in Fig.~\ref{fig:3}.

From Fig.~\ref{fig:3} one can conclude that values for $M_{1/2}$
$\lsim$ 1~TeV are favored by this observable, $M_{1/2}$ $\sim 500$~GeV giving
the best fit.

Finally we note that not all observables are practically independent
from $\lambda$: recall that within the present scenario, all sparticle
decays will proceed via the stau NLSP, since the couplings of the true
(singlino-like) LSP are of the order of $\lambda$ and hence small. Only
at the end of each MSSM-like decay chain, the stau NLSP will decay into
the singlino-like LSP, but its decay width can be tiny implying a
possibly visible stau track length \cite{singdecay}. We find that this
track length can be $\gsim 1$~mm at the LHC, if $\lambda \lsim 10^{-3}$;
this phenomenon can thus represent a possible ``smoking gun'' for the
cNMSSM.

\section{The NMSSM and GMSB}
Supersymmetric extensions of the SM with Gauge Mediated
Supersymmetry Breaking always involve messenger supermultiplets $\phi_i$
with a (supersymmetric) mass $M_{mess}$, but whose  CP-even and CP-odd
scalar masses squared are split by $m^2$. Possible origins of the SUSY
breaking parameter $m^2$ are
\begin{itemize}
\item Dynamical SUSY Breaking (non-perturbative) in a hidden sector
containing a SUSY Yang-Mills theory plus matter, and couplings of
$\phi_i$ to the hidden sector \cite{dsb};
\item O'Raifeartaigh-type models \cite{oraif};
\item models based on No-Scale supergravity \cite{noscale} with
Giu\-dice-Masiero-like terms \cite{gm} for $\phi_i$ in the K\"ahler
potential \cite{ue95}.
\end{itemize}

Since the messenger fields $\phi_i$ carry $SU(3)\times SU(2)\times
U(1)_Y$ gauge quantum numbers, they generate gaugino masses (at 1 loop)
and masses for all non-singlet scalars (at 2 loops) of the order
$M_{SUSY} \sim \frac{m^2}{16\pi^2 M_{mess}}$, but none of the
phenomenologically required $\mu$- or $B$-terms of the MSSM -- hence the
$\mu$-problem is even more pressing in general GMSB-like models.

Again, the simplest solution of the problem is the introduction of a
singlet $S$ together with its coupling $\lambda$ to $H_u$ and $H_d$
\footnote{The results of this section have been obtained in
collaboration with C.-C. Jean-Louis and A. M. Teixeira in
\cite{nmgmsb}.}.
However, soft SUSY breaking terms in the potential for the
singlet are necessary in order to generate a sufficiently large VEV of
$S$. In order to generate such terms radiatively (of the desired order),
it seems necessary to introduce a direct coupling $\sim \eta S\phi_i
\phi_i$ of $S$ to the messenger sector.

Then, integrating out the messengers generates desired terms like $m_S^2$
and $A_\lambda = \frac{1}{3} A_\kappa$; possibly, however, also terms
linear in $S$ in the superpotential $W \sim \xi_F S$ and in the
potential $V_{soft} \sim \xi_S S$, so-called ``tadpole terms''. Such
tadpole terms always trigger a non-vanishing $\left< S\right> \neq 0$
but, if allowed at 1 loop order, the radiatively generated parameters
$\xi_F$, $\xi_S$ tend to be somewhat large; one finds~\cite{ue95}
\beq\label{eq:xifxis}
\xi_F \sim \eta\, M_{mess}\, M_{SUSY},\quad \xi_S \sim 16\pi^2\,
\eta\, M_{mess}\, M_{SUSY}^2\; ,
\eeq
and recall that we typically expect $M_{mess} > M_{SUSY}$. On the other
hand, $\xi_S$ should not be larger than $M_{SUSY}^3$, which is the case
if $\eta \lsim \frac{M_{SUSY}}{16\pi^2 M_{mess}}$ typically implying
$\eta \lsim 10^{-5}$. 

\begin{figure}[!ht]
\includegraphics[width=5.5cm,angle=-90]{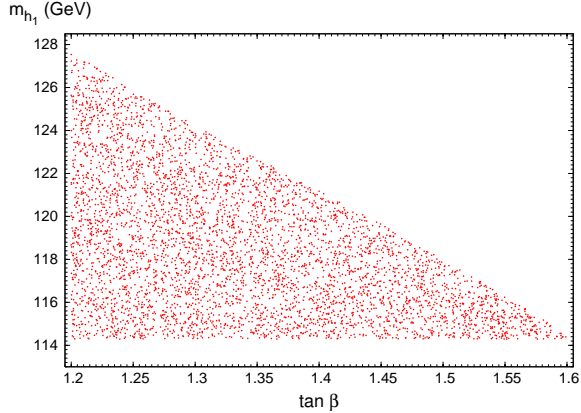}
\caption{$m_{h_1}$ as function of $\tan\beta$ in the scenario with
tadpole terms.}
\label{fig:4}
\end{figure}

As investigated in \cite{nmgmsb}, such models can be
phenomenologically viable, if $\lambda \gsim 0.5$ (and $\tan\beta \lsim
2$); then the NMSSM specific contribution $\sim \lambda^2$ to the scalar
Higgs mass matrix squared \cite{nmssm} pushes the lightest Higgs mass
$m_{h_1}$ above the LEP bound.
For the parameter choices $M_{mess}=10^6$ GeV and $M_{SUSY}=500$ GeV, we
have varied the parameters $0.5 < \lambda < 0.6$ and $10^{-6} < \eta <
10^{-5}$; the resulting values for $m_{h_1}$ are shown in
Fig.~\ref{fig:4} as function of $\tan\beta$.

The other Higgs states are heavier than $\sim 600$~GeV, the bino, wino
and slepton masses are in the range 110 to 290~GeV, and the squark and
gluino masses in the range 640 to 890~GeV; hence the entire Higgs and
sparticle spectrum satisfies all collider constraints for this class of
models inspite of the presence of tadpole terms for $S$.

Tadpole terms for $S$ can also be forbidden by discrete symmetries, if
the messenger sector is enlarged to $\phi_1$, $\overline{\phi}_1$,
$\phi_2$, $\overline{\phi}_2$ \cite{giuratt} and the superpotential is
chosen as
\beq\label{eq:gr}
W=\eta\,S\,\overline{\phi}_1 \phi_2 + M_{mess}(\overline{\phi}_1 \phi_1
+ \overline{\phi}_2 \phi_2) \; . 
\eeq

The soft terms $m_S^2$ ($<0$), $A_\kappa,\ A_\lambda$ are calculable in
terms of $\eta$ and $M_{SUSY}$ as before. Phenomenologically viable
regions in the parameter space $M_{SUSY}$, $M_{mess}$, $\eta$, $\lambda$
and $\tan\beta$ have been found in \cite{dgs} (and confirmed in 
\cite{nmgmsb}) where, however, the sparticle spectrum turns out to be
quite heavy: Bino, wino and slepton masses are in the range 450 to
1100~GeV, and the squark and gluino masses around 2~TeV.

In \cite{nmgmsb}, we have also investigated scenarios where the soft
terms $A_\kappa,\ A_\lambda$ are negligibly small at $M_{mess}$, i.e.
where all soft terms for the singlet vanish at $M_{mess}$ except for
$m_S^2$ (a corresponding hidden sector remains to be constructed). Then,
the scalar sector of the NMSSM has an R-symmetry (at $M_{mess}$), which
is, however, broken by radiative corrections to $A_\kappa$, $A_\lambda$
induced by the gaugino mass terms. Then, the explicit R-symmetry
breaking at the weak scale by $A_\lambda$, $A_\kappa \sim$ a~few GeV is
small (if $M_{mess}$ is not too large), and the spontaneous R-symmetry
breaking by $\left<H_u\right>$, $\left<H_d\right>$, $\left<S\right> \neq
0$ generates a pseudo Goldstone Boson, the lightest CP-odd Higgs scalar
$a_1$ \cite{raxion}. Consequently, the lightest Higgs scalar $h_1$ can
decay via $h_1 \to a_1 a_1$ escaping LEP constraints if $m_{h_1} \gsim
90$~GeV (depending on $m_{a_1}$) \cite{lhg}.

\begin{figure}[!ht]
\includegraphics[width=5.5cm,angle=-90]{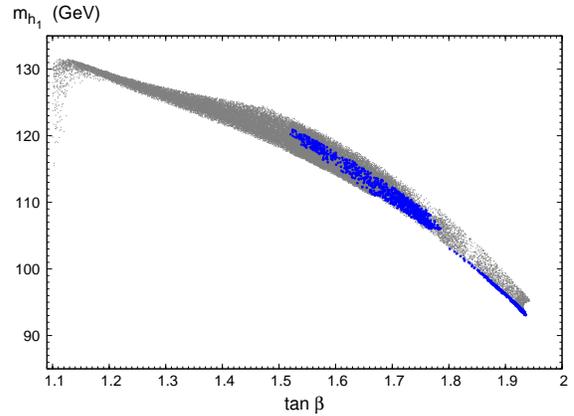}
\caption{$m_{h_1}$ as function of $\tan\beta$ in the scenario with
$A_\kappa,\ A_\lambda \sim 0$.}
\label{fig:5}
\end{figure}

We have studied phenomenologically viable regions in the parameter space
of such a scenario for $\lambda = 0.6$, $10^7$~GeV $< M_{mess}<5\cdot
10^9$~GeV and 200~GeV $< M_{SUSY}<280$~GeV as shown in Fig.~\ref{fig:5}
where, for $m_{h_1} < 114$~GeV, $m_{a_1}$ is below $m_{h_1}/2$.

Here the bino, wino and slepton masses are $\sim$~100 -- 200~GeV, the
squark and gluino masses $\sim$~450 -- 600~GeV, and the masses of the
additional Higgs bosons above $\sim 500$~GeV. The blue points satisfy
also the $2\sigma$ constraints on the muon anomalous magnetic moment.

Altogether a variety of NMSSM models with GMSB -- with and without
tadpole terms -- is phenomenologically viable, provided that the singlet
couples directly to the messengers such that destabilizing terms in the
singlet potential can be radiatively generated.

\section{Extra $U(1)'$ gauge symmetry}

A natural question is the one for a possible origin of a SM singlet
superfield like the $S$ of the NMSSM. In fact, multiplets of large GUT
gauge groups (like, e.g., $E_6$ \cite{king}) typically contain singlets
under the SM gauge groups which are, however, charged under one (or
more) extra $U(1)'$ gauge group(s) (see \cite{langack} for a recent
review). Quarks, leptons as well as the MSSM doublets $H_u$ and $H_d$
carry such $U(1)'$ charges as well, as a consequence of which the MSSM
$\mu H_u H_d$-term is forbidden and has to be generated by a VEV of $S$
(and a coupling $\lambda S H_u H_d$) as before.

Due to the $U(1)'$ charge of $S$, the $\kappa S^3$-term in the
superpotential of the NMSSM is forbidden as well, but the $S$-dependent
potential can still be stabilized for large $\left< S\right>$ due to the
$U(1)'$ -- D-term $\sim g'^2 |S|^4$. The $U(1)'$ -- D-term leads also to
additional $g'^2 |H_{u,d}|^4$-terms in the scalar potential, which imply
heavier (SM-like) physical Higgs scalars which satisfy more easily the
lower LEP bound of 114~GeV.

However, the cancellation of all anomalies (at scales $\sim M_{SUSY}$)
usually requires additional exotic matter (and possibly several SM
singlets) with masses of the order $M_{SUSY}$, as a consequence of which
the unification of the SM gauge couplings at $M_{GUT}$ is no longer
``automatic'' as in the MSSM or in the NMSSM.

The most evident phenomenological implication of such models is the
presence of at least one extra $Z'$ gauge boson; however, since it tends
to mix with the $Z$ boson of the SM, one obtains constraints on its mass
and the quantum numbers of matter whose loops are responsible for this
mixing. Also the neutralino sector is enlarged \cite{choi}, involving
extra states from both $Z'$- and SM singlet matter supermultiplets.

\section{Summary}

Under the assumption that the SUSY breaking scale $M_{SUSY}$ generates
the weak scale $\sim M_Z$, and no other dimensionful parameters are
present in the effective Lagrangian below the GUT scale, the NMSSM is
the most natural supersymmetric extension of the Standard Model.

If one adds the assumption of universal soft SUSY breaking terms
$M_{1/2}$, $m_0$ and $A_0$ one finds that the
phenomenologically viable range -- satisfying all present constraints
from collider- and $B$-physics as well as the dark matter relic density
-- for $M_{1/2}$, $m_0$ and $A_0$ in the cNMSSM is very different from
the cMSSM: it is caracterized by $m_0 \ll M_{1/2}$ and $A_0 \sim
\frac{1}{4} M_{1/2}$; the entire Higgs and sparticle spectrum can
finally be parametrized by $M_{1/2}$ only. The most notable feature of
this scenario is that the LSP is always singlino-like; depending on the
Yukawa coupling $\lambda$, a large NLSP (stau) lifetime can lead to
tracks of observable length at the end of sparticle decay chains at the
LHC.

In the framework of models with Gauge Mediated Supersymmetry Breaking,
the NMSSM allows to solve the $\mu$-problem in a phenomenologically
viable way, {\it provided $S$ couples directly to the messenger sector}.
Then, radiative corrections generate the soft SUSY breaking terms for
$S$, which are required for a sufficiently large VEV $\left< S\right>$.
Tadpole terms are {\it not} dangerous if the coupling $\eta$ of $S$ to
the messengers is sufficiently small. Different scenarios can be
realized implying different phenomenologies in the Higgs and sparticle
sectors; possible are, amongst others, light CP-odd scalars
(pseudo Goldstone Bosons), or light CP-even scalars with a large singlet
component.

Hopefully, we will know more about the scenario realized in nature
within a few years from now.

\begin{theacknowledgments}
It is a pleasure to thank the organisers of SUSY~08 for a very
inspiring and fruitful conference.

This talk is based on work in collaboration with A. Djouadi, C.-C.
Jean-Louis, F. Domingo and A.M. Teixeira. We acknowledge support from
the French ANR project PHYS@COL\&COS.
\end{theacknowledgments}

\end{document}